\documentclass[12pt,letter]{article}
\usepackage[utf8]{inputenc}
\usepackage{amsmath}
\usepackage{amsfonts}
\usepackage{amssymb}
\usepackage{graphicx, subfigure}
\usepackage{algorithmicx}
\usepackage{algorithm}
\usepackage{algpseudocode}
\usepackage{natbib, url}
\bibpunct{(}{)}{;}{a}{}{,}
\usepackage[left=2cm,right=2cm,top=2cm,bottom=2cm]{geometry}

\author{W. James Murdoch and Mu Zhu}
\date{}
\title{Expanded Alternating Optimization\\of Nonconvex Functions with Applications\\to Matrix Factorization and Penalized Regression}
\begin{document}
\maketitle 

\begin{abstract}%
We propose a general technique for improving alternating optimization (AO) of nonconvex functions. Starting from the solution given by AO, we conduct another sequence of searches over subspaces that are both meaningful to the optimization problem at hand and different from those used by AO. To demonstrate the utility of our approach, we apply it to the matrix factorization (MF) algorithm for recommender systems and the coordinate descent algorithm for penalized regression (PR), and show meaningful improvements using both real-world (for MF) and simulated (for PR) data sets. Moreover, we demonstrate for MF that, by constructing search spaces customized to the given data set, we can significantly increase the convergence rate of our technique. 
\end{abstract}
\vspace{5mm}
\newpage

\newcommand{\myvec}[1]{\boldsymbol{#1}}
\def\T{{\mbox{\rm\tiny T}}}
\def\ao{\mathcal{S}_{AO}^{(t)}}
\def\es{\mathcal{S}_{escape}^{(t)}}

\begin{section}{Introduction}
\label{sec:intro}

Alternating optimization (AO) is a commonly used technique for finding the extremum of a multivariate function, $f(\myvec{z})$, where $\myvec{z} \in \mathbb{R}^d$. In this approach, one breaks up the (multi-dimensional) input variable $\myvec{z}$ into a few blocks, say, $\myvec{z}_1, \myvec{z}_2, ..., \myvec{z}_B$, and successively optimizes the objective function over each block of variables while holding all other blocks fixed. That is, one solves  
\begin{eqnarray}
\underset{\myvec{z}_b}{\min} \quad f(\myvec{z}_1,\myvec{z}_2,...,\myvec{z}_B) \label{eq:indivprob}
\end{eqnarray}
successively over $b=1,2,...,B,1,2,...,B,...$ until convergence is achieved. This is an especially natural approach when each individual optimization problem (\ref{eq:indivprob}) over $\myvec{z}_b$ is relatively easy to solve. Two well-known examples in statistics are: matrix factorization and penalized regression, but there are many others. 

\subsection{Matrix factorization}
\label{sec:MF}

The Netflix contest drew much attention to the matrix factorization problem \citep{Koren09, netflix-review-statsci, zhu-ssc14}. Given a user-item rating matrix $\myvec{R}$, whose element $r_{ui}$ is the rating of item $i$ by user $u$, the goal is to find low-rank matrices $\myvec{P}$ and $\myvec{Q}$, such that
\[
 \myvec{R} \approx \myvec{P}\myvec{Q}^{\T} = 
 \underbrace{%
 \left[
 \begin{array}{c}
 \myvec{p}_1^{\T} \\
 \myvec{p}_2^{\T} \\
 \vdots \\
 \myvec{p}_n^{\T} 
 \end{array}
 \right]}_{n \times K}
 \underbrace{%
 \left[
 \begin{array}{cccc}
 \myvec{q}_1 & \myvec{q}_2 & \cdots & \myvec{q}_m 
 \end{array}
 \right]}_{K \times m}.
\]
The vector $\myvec{p}_u$ can be viewed as the coordinate of user $u$ in a $K$-dimensional map and the vector $\myvec{q}_i$, the coordinate of item $i$. With these coordinates, it is then possible to recommend item $i$ to user $u$ if $\myvec{q}_i$ and $\myvec{p}_u$ are closely aligned. 
Since we don't know every user's preferences on every item, many entries of $\myvec{R}$ are missing. Let
\[
T = \{(u,i): \quad \mbox{$r_{ui}$ is known}\}
\]
be the set of observed ratings. In order to estimate these user- and item-coordinates, we can solve the following optimization problem:
\begin{eqnarray}
\label{eq:MF}
 \underset{\myvec{P},\myvec{Q}}{\min} \quad 
 L(\myvec{P},\myvec{Q}) \equiv
 \sum_{(u,i) \in T} \left(\myvec{r}_{ui} - \myvec{p}_u^{\T} \myvec{q}_i\right)^2 +
 \lambda \left(\sum_{u=1}^n \|\myvec{p}_u\|^2+ \eta\sum_{i=1}^m \|\myvec{q}_i\|^2\right)
\end{eqnarray}
where the bracketed terms being multiplied by $\lambda > 0$ are penalties on the parameters being estimated, introduced to avoid over-fitting, because $n$ and $m$ are typically quite large relative to the number of known ratings (or the size of the set $T$). Here, we follow the work of \citet{zhu-sam13} and use an extra factor $\eta =n/m$ to balance the penalties imposed on the two matrices, $\myvec{P}$ and $\myvec{Q}$. 

It is natural to use AO for solving (\ref{eq:MF}). With both $\myvec{p}_u$ and $\myvec{q}_i$ being unknown, (\ref{eq:MF}) is not a convex problem, but once we fix all $\myvec{p}_{v}$ ($v \neq u$) and $\myvec{q}_i$, the individual problem
\[
 \underset{\myvec{p}_u}{\min} \quad L(\myvec{P},\myvec{Q})
\]
over $\myvec{p}_u$ is convex and hence easy to solve.

\subsection{Penalized regression}
\label{sec:PR}

During the last decade, penalized regression techniques have attracted much attention in the statistics literature \citep{lasso, scad, mcp}. Suppose that $\myvec{y}, \myvec{x}_1, ..., \myvec{x}_d \in \mathbb{R}^n$ are all properly standardized to have mean zero ($\myvec{1}^{\T}\myvec{y}=0$, $\myvec{1}^{\T}\myvec{x}_j=0$) and variance one ($\|\myvec{y}\|=1$, $\|\myvec{x}_j\|=1$). The prototypical problem can be expressed as follows:
\begin{eqnarray}
\label{eq:PR}
 \underset{\myvec{\beta}}{\min} \quad 
 L(\myvec{\beta}) \equiv \|\myvec{y} - 
 (\beta_1 \myvec{x}_1 + \beta_2 \myvec{x}_2 + \cdots + \beta_d \myvec{x}_d)
 \|^2 + \sum_{j=1}^d J(\beta_j),
\end{eqnarray}	
where $J(\cdot)$ is a penalty function. Many different penalty functions have been proposed. A widely used class of penalty functions is  
\[
 J(\beta_j) = \lambda |\beta_j|^{\alpha}.
\]
The case of $\alpha=2$ is known as the ridge penalty \citep{ridge},
and that of $\alpha=1$ is known as the LASSO \citep{lasso}. In both of these cases, the function $J(\cdot)$ is convex. 
In recent years, \emph{nonconvex} penalty functions have started to garner the attention of the research community, e.g., the SCAD \citep{scad}, and the MC+ \citep{mcp}:
\begin{eqnarray}
\label{eq:mcp}
 J(\beta_j) = \lambda
 \int_{0}^{|\beta_j|} \left(1-\frac{x}{\gamma\lambda}\right)_+ dx = 
 \begin{cases}
 \lambda |\beta_j| - \frac{\beta_j^2}{2\gamma}, &  |\beta_j| \leq \gamma \lambda; \\
 \frac{1}{2} \gamma \lambda^2,                  &  |\beta_j| > \gamma \lambda.
 \end{cases}
\end{eqnarray}
We will focus on the MC+ in this paper; hence, details of the SCAD are omitted, and we refer the readers to the original papers \citep{scad,mcp} for explanations of for why these particular types of penalty functions are interesting.  

Currently, the preferred algorithm for fitting these penalized regression models is the coordinate descent algorithm \citep{jht-cd}. One can view the coordinate descent algorithm as the ``ultimate'' AO strategy. For given $\lambda>0$, the coordinate descent algorithm solves
\[
\underset{\beta_j}{\min} \quad L(\myvec{\beta}),
\]
while fixing all $\beta_k$ ($k \neq j$), successively over $j=1,2,...,d,1,2,...d,...$ until convergence. In fact, sometimes the general AO algorithm, with which we started this article, is dubbed the ``blockwise coordinate descent'' algorithm.

\subsection{Saddle points}
\label{sec:saddleprob}

For the ridge penalty and the LASSO, the penalty function $J(\cdot)$ is convex, so the coordinate descent algorithm behaves ``well''. But for nonconvex penalty functions such as the SCAD and the MC+, there is no guarantee that the coordinate descent algorithm can reach the global solution of (\ref{eq:PR}). Exactly the same point can be made about the AO algorithm for solving (\ref{eq:MF}). 

In fact, for these nonconvex problems, not only can the AO strategy get stuck at inferior local solutions, but it also can be trapped at saddle points. \citet{benjio-saddle} found that getting stuck at a saddle point can be a far more serious problem than getting trapped at a local minimum. \citet{tayal} proposed an intriguing method to improve AO by facilitating AO algorithms to escape saddle points. They introduced the concept of a shared ``perspective variable'' 
(more details in Section~\ref{sec:scale}), but lacked intuition for why this was a good idea. 

\subsection{Our contribution}

In this article, we begin by interpreting the approach of \citet{tayal} geometrically --- in particular, sharing a ``perspective variable'' results in an expanded search space at each step. However, searching over a slightly larger space at each step comes with a higher computational cost, which should be avoided if possible. Thus, our proposal is as follows: first, run the faster AO iterations until convergence; then, try to escape being trapped at an undesirable location by searching over a different space. Of critical importance is the choice of the search space. To this end, we introduce the important idea of defining search spaces that depend upon the particular data observed, as opposed to traditional techniques, including AO and those in \citet{tayal}, that fix the search spaces a priori. We apply these ideas to improve the AO algorithm for solving (\ref{eq:MF}) as well as the coordinate descent algorithm for solving (\ref{eq:PR}), with a focus on the MC+ penalty. However, we stress that this is a general algorithm that can be applied to other AO problems as well. 

\subsection{Notation}
	
In what follows, we will use the notation $\myvec{z}_{-b}$ to denote all other components except those in block $b$.

\end{section}

\begin{section}{Motivation}
\label{sec:motivation}

It is convenient for us to motivate our key ideas with a very simple example. 

\begin{subsection}{A simple example}
\label{sec:simple-ex}

Consider the function, 
\[
f(x,y) = (x-y)^2 - x^2 y^2.
\]
Suppose that we attempt to minimize $f(x,y)$ with AO, and that, at iteration $t$, we have reached the point $(x_t,y_t) = (0,0)$. While fixing $x_t=0$, $f(0,y)=y^2$ is minimized at $y=0$. Similarly, $x=0$ is the optimal point when $y_t$ is fixed at 0. Thus, the AO algorithm is stuck at $f(0,0)=0$. However, it is easy to see that $f(x,y)$ is actually unbounded below, and that $(0,0)$ is a saddle point. 

At $(x_t,y_t)=(0,0)$, the search space defined by AO is 
\begin{eqnarray*}
\ao &=& \{(x,y): x=0\} \cup \{(x,y): y=0\}.
\end{eqnarray*} 
In this case, we can see that being restricted to this particular search space is the very reason why we are trapped at the saddle point. Therefore, if we could use a slightly different search space, we might be able to escape this saddle point. For example, Figure~\ref{fig:toy} shows that using the search space,
\begin{eqnarray*}
\es &=& \{(x,y): x=y\},
\end{eqnarray*} 
would suffice.

\begin{figure}[hbtp]
\centering
\includegraphics[width=0.475\textwidth]{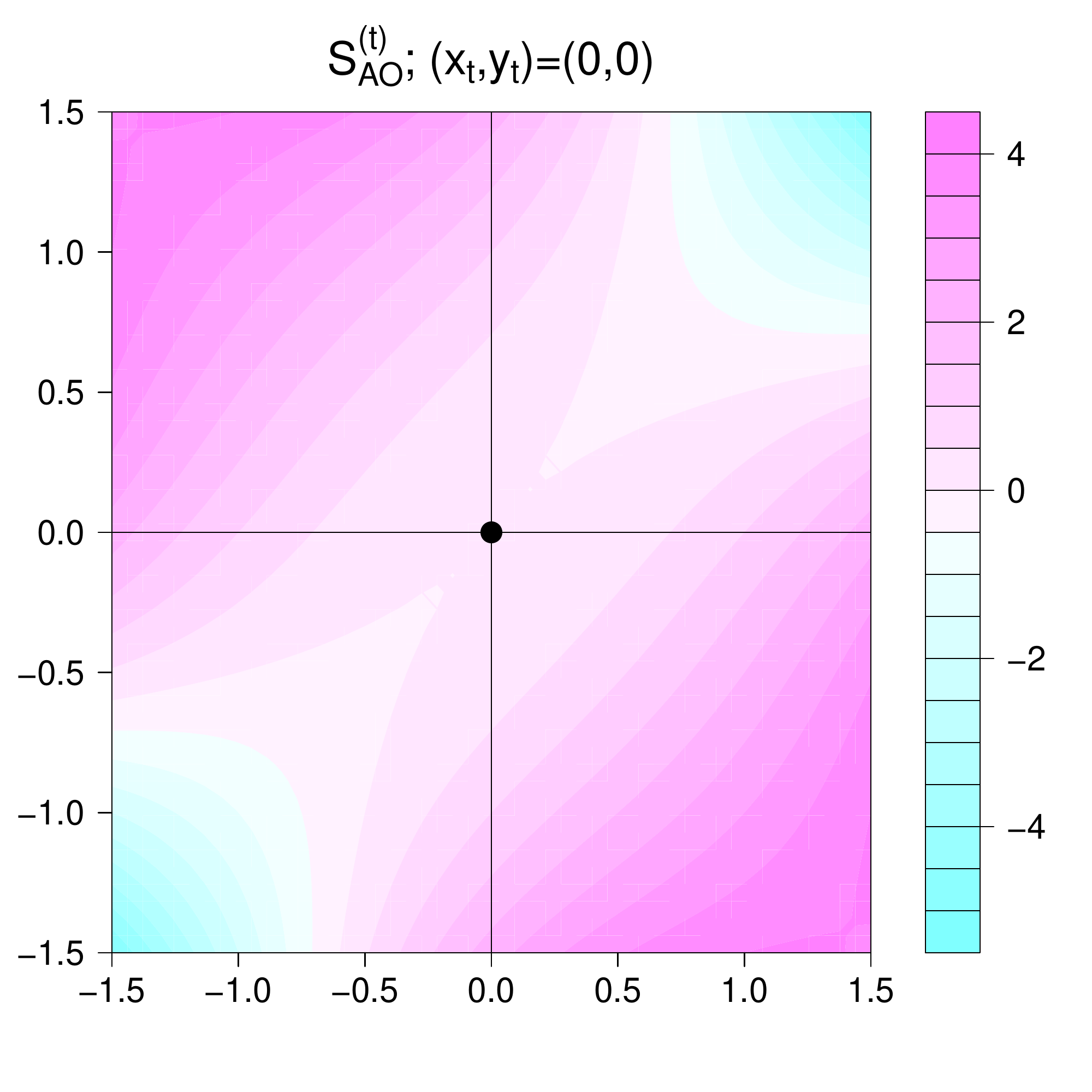}
\includegraphics[width=0.475\textwidth]{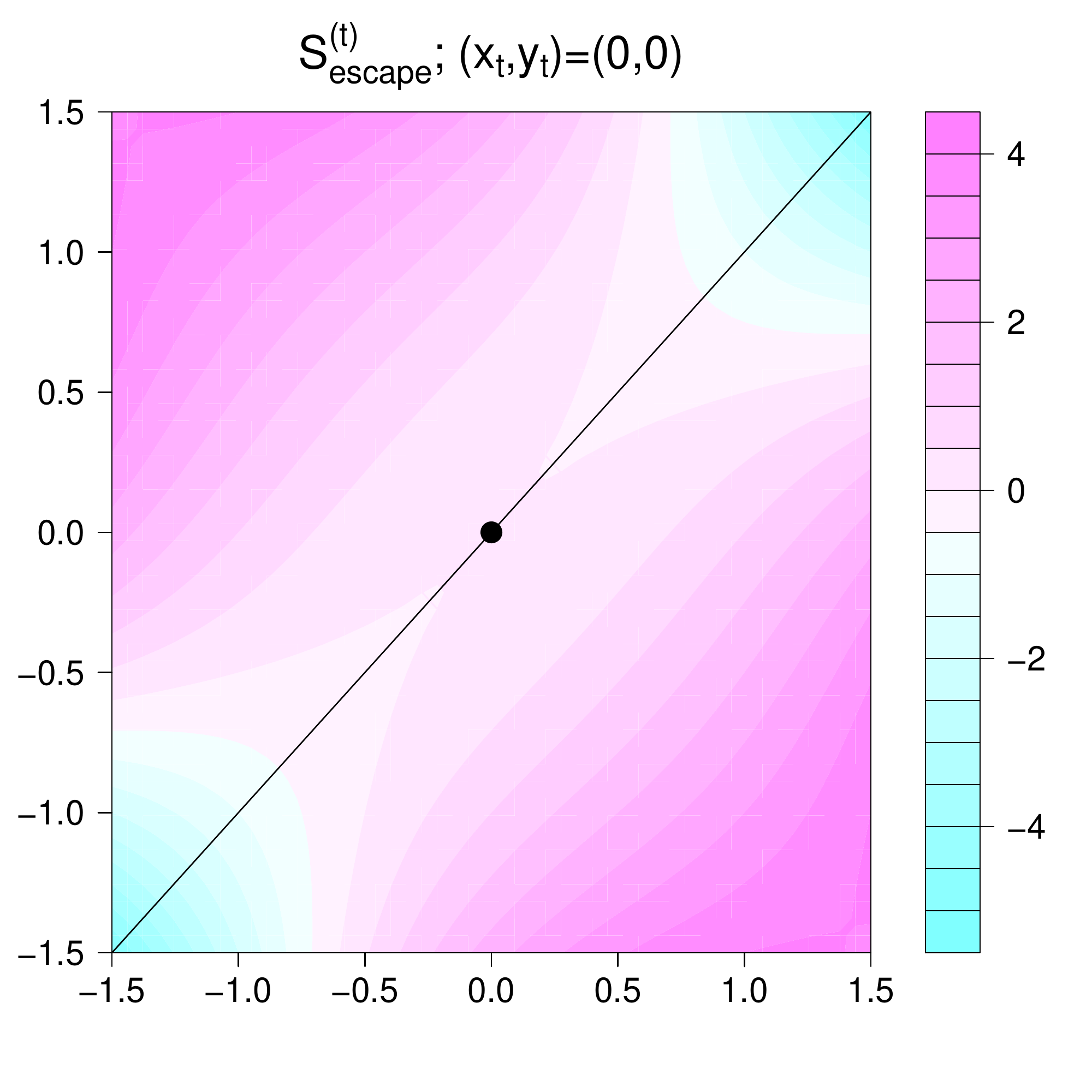}
\caption{Different search spaces, $\ao$ and $\es$, at $(x_t,y_t)=(0,0)$, superimposed on a contour plot of $f(x,y)=(x-y)^2-x^2 y^2$.}
\label{fig:toy}
\end{figure}

\end{subsection}

\begin{subsection}{Main idea}

The main lesson from the simple example above is that the restricted search space defined by the AO strategy, $\mathcal{S}_{AO}$, may cause the search to be trapped at undesirable locations such as saddle points, and that we may escape such traps by conducting the search in a slightly different space, $\mathcal{S}_{escape}$. This observation naturally leads us to propose the following strategy. 

First, we run standard AO --- that is, search in $\mathcal{S}_{AO}^{(1)}$, $\mathcal{S}_{AO}^{(2)}$, ..., $\mathcal{S}_{AO}^{(\tau)}$ --- until convergence. Then, starting from the AO solution, we continue searching in a different sequence of spaces --- $\mathcal{S}_{escape}^{(\tau+1)}$, $\mathcal{S}_{escape}^{(\tau+2)}$, ..., $\mathcal{S}_{escape}^{(\tau+\tau')}$ --- until convergence. If we see a sufficient amount of improvement in the objective function --- an indication that the escaping strategy ``worked'', then we repeat the entire process using the improved result as the new starting point; otherwise, the algorithm terminates (see Algorithm~\ref{algo:main}). 

Needless to say, the key to the strategy we outlined above lies in the definition of the escaping sequence, $\mathcal{S}_{escape}^{(t)}$. We discuss this next, in Section~\ref{sec:escape}.

\begin{algorithm}
\caption{\label{algo:main}%
A General Form of our Proposed Technique for Minimizing $f(\cdot)$}
\begin{algorithmic}
\Function{Escape}{Start}
\State $\text{Result} \gets \text{standardAO}(\text{Start})$ 
\State $\text{newResult} \gets \text{SearchOverDifferentSpaces}(\text{Result})$
\If {$f(\text{newResult}) < f(\text{Result}) - \varepsilon$}
	\State return(Escape(newResult))
\Else
	\State return(newResult)
\EndIf
\EndFunction
\end{algorithmic}
\end{algorithm}

\end{subsection}

\section{Escaping strategies}
\label{sec:escape}

The key insight from Section~\ref{sec:simple-ex} is that switching to a different search space than the one defined by AO may allow us to escape being trapped at undesirable locations. In this section, we describe how these different search spaces can be specified.  

\subsection{Scaling}
\label{sec:scale}

The proposal by \citet{tayal} of sharing a ``perspective variable'', which we alluded to earlier in Section~\ref{sec:saddleprob}, essentially amounts to the following. At each alternating step, rather than solving (\ref{eq:indivprob}), they proposed that we solve  
\begin{eqnarray}
\label{eq:scaledstep}
\underset{\myvec{z}_b, v_b}{\min} &\quad& 
 f\left(%
   v_b \myvec{z}_1, ..., v_b \myvec{z}_{b-1}, 
   ~ \myvec{z}_b, 
   ~ v_b \myvec{z}_{b+1}, ..., v_b \myvec{z}_B 
   \right)
\end{eqnarray} 
instead, where $v_b \in \mathbb{R}$ is the so-called ``perspective variable''. That is, we no longer just search for the optimal $\myvec{z}_b$ while keeping $\myvec{z}_{-b}$ fixed. When searching for the optimal component $\myvec{z}_b$, we are free to \emph{scale} all other components as well. Suppose the optimal scaling variable coming out of solving (\ref{eq:scaledstep}) is $v_b^*$. The component $\myvec{z}_{-b}$ is then adjusted accordingly, i.e.,
\begin{eqnarray*}
 \myvec{z}_{-b} &\longleftarrow& v_b^{*} \myvec{z}_{-b},
\end{eqnarray*}
before the next alternating step (for optimizing over $\myvec{z}_{b+1}$ and scaling $\myvec{z}_{-(b+1)}$) begins.

When so described, it is somewhat mysterious why it helps to scale $\myvec{z}_{-b}$ when solving for $\myvec{z}_b$. However, when viewed in terms of their respective search spaces, we can interpret this proposal as one particular way to define the search space $\mathcal{S}_{escape}$. 

As illustrated in Figure~\ref{fig:showspace}, at iteration $t$, the search space defined by AO is 
\[
\mathcal{S}_{AO}^{(t)} = \{(\myvec{z}_b,\myvec{z}_{-b}): \myvec{z}_{-b}=\myvec{z}_{-b}^{(t-1)}\} = 
\myvec{z}^{(t-1)} + \mbox{span}\{\myvec{z}_{b}\};
\]
whereas, if we are free to scale $\myvec{z}_{-b}$ at the same time, the search space becomes
\begin{eqnarray*}
\mathcal{S}_{escape}^{(t)} 
&=& \{(\myvec{z}_b, \myvec{z}_{-b}): \exists~v \in \mathbb{R} \mbox{ such that } \myvec{z}_{-b} = v \myvec{z}_{-b}^{(t-1)} \} \\
&=& \{ \lambda \myvec{z}_{-b}^{(t-1)} + \myvec{x} | \lambda \in \mathbb{R}, \myvec{x} \in \mbox{span}\{\myvec{z}_b\} \}.
\end{eqnarray*}
Clearly, $\mathcal{S}_{escape}^{(t)}$ is larger than $\mathcal{S}_{AO}^{(t)}$ (but still much smaller than the entire space $\mathbb{R}^d$). Thus, one way to understand the idea of freely scaling $\myvec{z}_{-b}$ while optimizing over $\myvec{z}_b$ is that it allows us to conduct our search in a slightly larger subspace, thereby improving our chances of finding a better solution. 

\begin{figure}[h]
\centering
\includegraphics[width=0.75\textwidth]{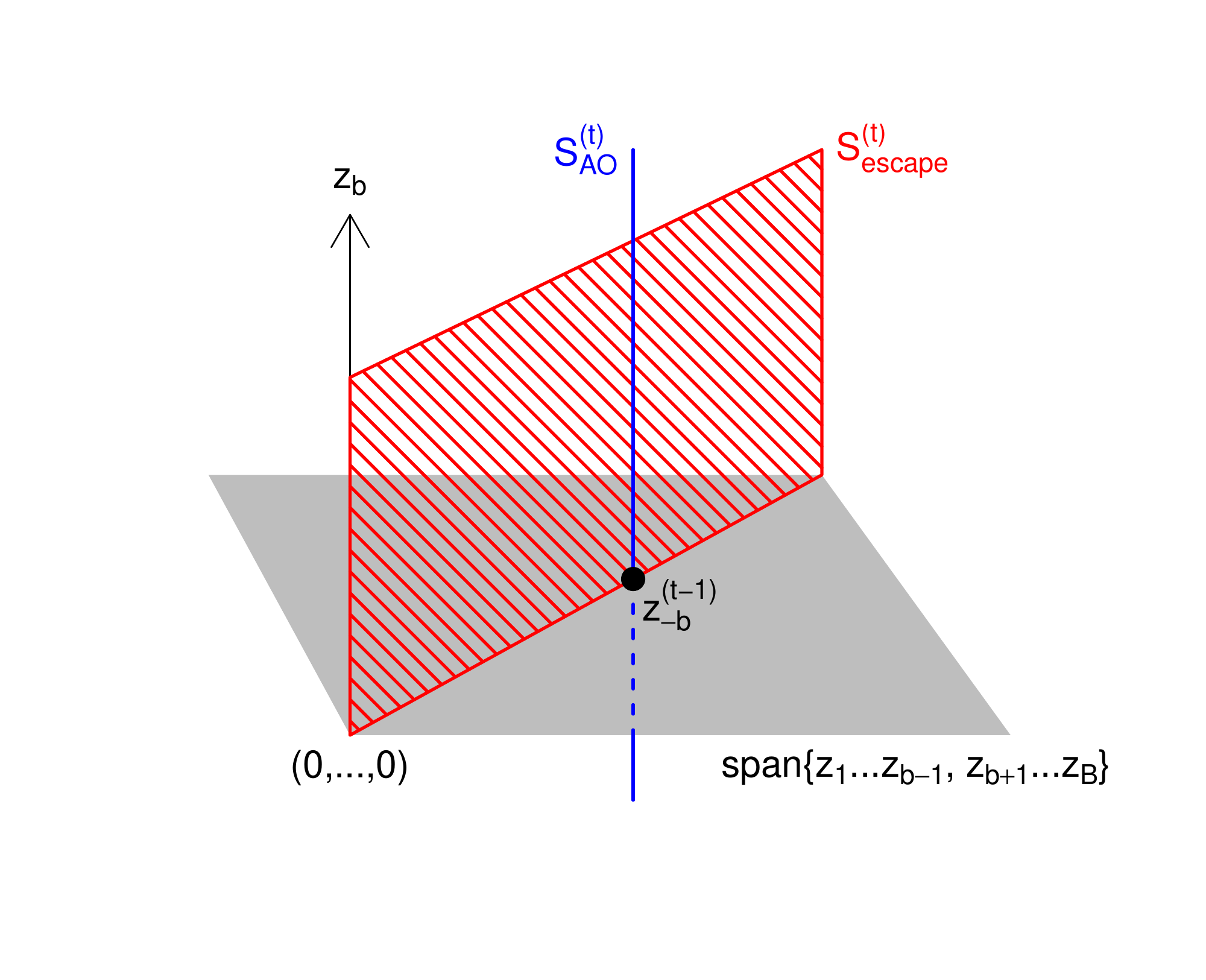}
\vspace{-1cm}
\caption{\label{fig:showspace}%
Illustration of $\mathcal{S}_{AO}^{(t)}$ versus $\mathcal{S}_{escape}^{(t)}$ as implied by the idea of sharing a ``perspective variable'' --- equation~(\ref{eq:scaledstep}). }
\end{figure}

\subsection{Restricted joint search}
\label{sec:joint}

This particular point of view immediately suggests that there are many other ways to expand, or simply alter, the search space. For example, once the AO steps have converged, we can try to escape by solving a \emph{restricted} joint optimization problem such as  
\begin{eqnarray}
\label{eq:linearjoint}
\underset{\alpha_1,...,\alpha_B}{\min} &\quad& 
 f\left(%
   \myvec{z}_1+\alpha_1 \myvec{w}_1 I_1, 
   \myvec{z}_2+\alpha_2 \myvec{w}_2 I_2, ...,
   \myvec{z}_B+\alpha_B \myvec{w}_B I_B \right), 
\end{eqnarray} 
where 
\[
\myvec{w}_b \in 
\mbox{span}\{\myvec{z}_b\}, 
\quad b = 1, 2, ..., B,
\] 
are some pre-chosen directions (more about these later), and
\[
I_b = 
\begin{cases}
1, & \mbox{if component $b$ is chosen to participate in this restricted joint optimization step}; \\
0, & \mbox{otherwise}.
\end{cases}
\]
The kind of search spaces generated by (\ref{eq:linearjoint}) can be described as
\[
\mathcal{S}_{escape}^{(t)} = \myvec{z}^{(t-1)}+\mbox{span}\{\myvec{w}_b: I_b \neq 0\};
\]
see Figure~\ref{fig:showspace2} for an illustration. The restricted joint search problem (\ref{eq:linearjoint}) can be viewed as a compromise between using a different search space --- i.e., $\es$ rather than $\ao$ --- and avoiding a full-scale, simultaneous search over the entire space $\mathbb{R}^d$. 

\begin{figure}[h]
\centering
\includegraphics[width=0.75\textwidth]{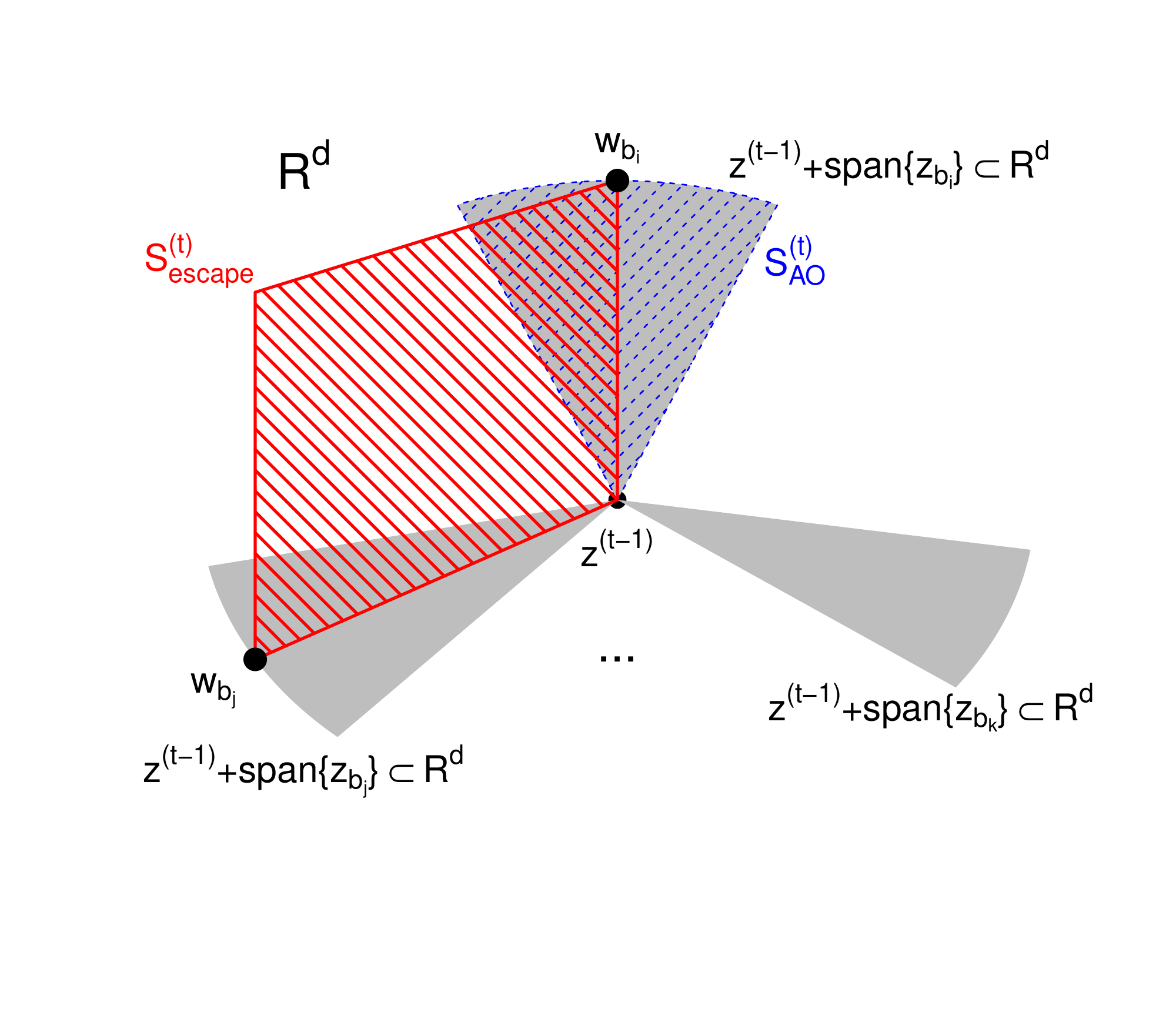}
\vspace{-1cm}
\caption{\label{fig:showspace2}%
Illustration of $\mathcal{S}_{AO}^{(t)}$ versus $\mathcal{S}_{escape}^{(t)}$ as implied by the restricted joint search problem (\ref{eq:linearjoint}). The three shaded areas denote three different subspaces of $\mathbb{R}^d$. The  ellipsis ($\cdots$) denotes the fact that many other subspaces are not shown. One of these subspaces would correspond to $\mathcal{S}_{AO}^{(t)}$, e.g., $\myvec{z}^{(t-1)}+\mbox{span}\{\myvec{z}_{b_i}\}$. Here, $I_{b_i} = I_{b_j} = 1$ and $I_{b} = 0$ for all $b \neq b_i, b_j$ including $b_k$.}
\end{figure}

\end{section}

\begin{section}{Improved AO for matrix factorization}

In this section, we apply our escaping strategies to the matrix factorization problem described in Section~\ref{sec:MF}. In particular, after solving the optimization problem (\ref{eq:MF}) with AO, we switch to search over a different space so as to escape saddle points, and/or inferior local solutions.

\begin{subsection}{Scaling}
\label{sec:MFscale}

As we described in Section~\ref{sec:scale}, introducing a shared variable allows us to search over a slightly expanded space. For fixed $\myvec{Q}$, this strategy minimizes
\begin{align}
\label{eq:MFscale}
L(\myvec{P},v) = \sum_{u,i \in T} \left[r_{ui} - \myvec{p}_u^{\T} 
(v\myvec{q}_i) 
\right]^2 + 
\lambda \left(\sum_{u=1}^n \| \myvec{p}_u \|^2 + 
\eta \sum_{i=1}^m \left\| v\myvec{q}_i \right\|^2 \right)
\end{align}
over $\myvec{P}$ and $v$ simultaneously, which we solve using a quasi-Newton algorithm with BFGS updates \citep[see, e.g.,][]{pracopt}. Analogously, for fixed $\myvec{P}$, we also numerically optimize over $\myvec{Q}$ and a scaling variable $u$ for $\myvec{P}$. 

\end{subsection}

\begin{subsection}{Restricted joint search}
\label{sec:MFjoint}

As the loss function for matrix factorization is generally quite high-dimensional, we expect that only increasing the dimensionality of the search space by one can have only a limited effect. In addition, expanding the search space by introducing a scaling variable also limits the types of subspaces we can search over. These are the reasons why we proposed the restricted joint search problem (\ref{eq:linearjoint}) in Section~\ref{sec:joint}. 

For the matrix factorization problem, this proposal amounts to constructing a family of search vectors $\{\myvec{w}_p^i\}_{i=1}^n$ and $\{\myvec{w}_q^i\}_{i=1}^m$ corresponding to each user- and item-vector, respectively, and searching over all of these directions simultaneously. Mathematically, this corresponds to solving
\begin{multline}
\label{eq:MFjoint}
\underset{\myvec{\alpha},\myvec{\beta}}{\min}\quad
L(\myvec{\alpha},\myvec{\beta}) = \sum_{u,i \in T} 
\left[r_{ui} - 
\left(\myvec{p}_u + \alpha_u \myvec{w}_p^u\right)^{\T}
\left(\myvec{q}_i + \beta_i \myvec{w}_q^i\right)
\right]^2 + \\
\lambda \left(\sum_{u=1}^n \| \myvec{p}_u + \alpha_u \myvec{w}_p^u \|^2 + \eta \sum_{i=1}^m \| \myvec{q}_i + \beta_i \myvec{w}_q^i \|^2 \right),
\end{multline}
over $\myvec{\alpha} \in \mathbb{R}^n$ and $\myvec{\beta} \in \mathbb{R}^m$.
To make this approach computationally feasible for larger problems, we generally require that $\myvec{w}_q^i, \myvec{w}_p^u = \myvec{0}$ for all but a relatively small number of indices $u, i$. To do so, we sample each $u$ with probability $s/n$ and each $i$ with probability $s/m$, for some $s \ll n, m$. That is, on average, we randomly choose $s$ user-vectors and item-vectors to participate in the restricted joint optimization (\ref{eq:MFjoint}).  

Having established a framework for our desired search space, the most important remaining question is: how to choose our search vectors $\myvec{w}_q^i, \myvec{w}_p^u$? This requires a notion of what an informative subspace is to search over. Below, we describe two different approaches.

\begin{subsubsection}{Random choices of $\myvec{w}_p^u$ and $\myvec{w}_q^i$}
\label{sec:MFrand}

Trying to determine what set of vectors $\myvec{w}_p^u, \myvec{w}_q^i$ will produce the largest decrease in the loss function (\ref{eq:MFjoint}) is a challenging task. As such, it is a good idea to establish a simple baseline against which to measure more sophisticated spaces. Having such a baseline space will also serve to illustrate the power of our method in its simplest form. 

For our baseline, we simply choose our search vectors at random. Specifically, for each chosen index $u$, we sample $\myvec{w}_p^u \sim N(\myvec{0},\myvec{I})$ from the $K$-variate standard normal distribution, and likewise for each chosen index $i$. This procedure incorporates no information about the optimization problem at hand, nor the data. But, as we shall report below (Section~\ref{sec:rsltMF}), even these random choices of directions can lead to better solutions. Next, we provide a more sophisticated subspace that uses such information to produce better results.

\end{subsubsection}

\begin{subsubsection}{Greedy choices of $\myvec{w}_p^u$ and $\myvec{w}_q^i$}
\label{sec:MFopt}

In general, the optimal search space for a given loss function would depend upon specific properties of the function itself. However, none of our previous approaches (Section~\ref{sec:MFscale} and Section~\ref{sec:MFrand}) explicitly took such information into account. We now describe an approach that does. 

Suppose that, for $\myvec{p}_u$, we are searching for the optimal step size $\alpha$ in a given direction $\myvec{w}$, while keeping everything else fixed. The objective function for this particular search operation is 
\begin{eqnarray}
\label{eq:MFopt}
L(\alpha) = 
\sum_{i \in T_{u}}
\left[r_{u i} - \left(\myvec{p}_u + \alpha \myvec{w}\right)^{\T} \myvec{q}_i\right]^2 + 
\lambda \|\myvec{p}_u + \alpha \myvec{w}\|^2 + (\mbox{terms not depending on $\alpha$}),
\end{eqnarray}
where $T_u = \{i: r_{ui} \mbox{ is known}\}$. Differentiating (\ref{eq:MFopt}) with respect to $\alpha$ and setting it equal to zero, we can solve for the optimal $\alpha$ as a function of $\myvec{w}$:
\begin{eqnarray*}
\widehat{\alpha}(\myvec{w}) = 
\frac{\sum_{i \in T_{u}} \myvec{w}^{\T}\myvec{q}_i (r_{ui} - \myvec{p}_u^{\T}\myvec{q}_i) - 
      \lambda   
      \myvec{w}^{\T}\myvec{p}_u}
     {\sum_{i \in T_{u}}(\myvec{w}^{\T}\myvec{q}_i)^2 + 
	 \lambda 
	 \|\myvec{w}\|^2}.
\end{eqnarray*}
By plugging in the optimal step size $\widehat{\alpha}(\myvec{w})$ into (\ref{eq:MFopt}), the function $L(\alpha)$ becomes a function of $\myvec{w}$, $L(\widehat{\alpha}(\myvec{w}))$. We can now solve for the optimal search direction $\myvec{w}$, using standard numerical optimization techniques --- again, we use quasi-Newton with BFGS updates. 

In doing so, we have solved for a search direction $\myvec{w}$ such that letting $\myvec{p}_u$ take an optimal step in its direction will produce the maximal decrease in the overall loss function. We construct our set of search vectors $\{\myvec{w}_p^u\}$ by repeating this process for each chosen $\myvec{p}_u$, and the set of search vectors $\{\myvec{w}_q^i\}$ is obtained in the same fashion.

\end{subsubsection}

\end{subsection}

\end{section}

\begin{section}{Improved AO for MC+ regression}

In this section, we apply our escaping strategies to the penalized regression problem described in Section~\ref{sec:PR}. We focus on the MC+ penalty function, but our strategies can be applied to other nonconvex penalty functions as well. We also investigate the application of our method to fitting an entire regularization surface, as introduced by \cite{sparsenet}. Although the singularity of the MC+ penalty function $J(\cdot)$ at $0$ places certain limits on the types of subspaces that can be feasibly optimized over, applying our ideas in their simple forms still produces notable improvements. 

\begin{subsection}{Scaling}
\label{sec:MCPbasic}

Again, the idea of using a shared variable (Section~\ref{sec:scale}) applies. In this setting, this amounts to taking a number of expanded coordinate descent steps (after the standard coordinate descent steps steps have converged; see Algorithm~\ref{algo:main}), so that we simultaneously search over a single coefficient $\beta_j$, as well as a scaling variable for the rest of the coefficient vector, $\myvec{\beta}_{-j}$. Mathematically, these expanded coordinate descent steps solve 
\begin{eqnarray}
\label{eq:MCP-all}
\underset{\beta_j, v}{\min}\quad 
L(\beta_j,v) = 
\left\|\myvec{y} - \beta_j \myvec{x}_j - \sum_{k \neq j} (v\beta_k)\myvec{x}_k\right\|^2 +
J(\beta_j) + \sum_{k \neq j} J\left(v \beta_k\right),
\end{eqnarray}
for $j=1,2,...,d$.

We can actually solve for the optimal $\beta_j$ and $v$ explicitly in this case (see Appendix~\ref{sec:MCPdetail}). This is attractive because it allows us to avoid using numerical optimization for these steps, which would have been more difficult due to the non-differentiability of the penalty function at zero. The technical details for these steps are provided in the Appendix.

\end{subsection}

\begin{subsection}{Selective scaling}
\label{sec:MCPopt}

Using our general notation (Sections~\ref{sec:intro}--\ref{sec:escape}), coordinate descent corresponds to each ``block'' $\myvec{z}_b$ being one-dimensional. This puts a certain limit on the kind of restricted joint search operations (Section~\ref{sec:joint}) we can implement. In particular, for any given $\myvec{z}_b$, the only available choice of $\myvec{w}_b$ is $\myvec{z}_b$ itself. However, we are still free to determine which $I_b=1$. 

As we pointed out in Section~\ref{sec:MFopt}, tailoring the search space to the observed data can yield improved results. In the context of regression, it is useful to consider how changing the coefficient in front of $\myvec{x}_j$ could affect the coefficient in front of another variable, say $\myvec{x}_k$. If these two variables are independent, then a change in $\beta_j$ would not result in a change to the optimal $\beta_k$. However, if these two variables are highly correlated, then one would expect that a decrease in $\beta_j$ could lead to an increase or decrease in $\beta_k$ depending on whether their correlation is positive or negative, as some of the dependence previously accounted for by $\myvec{x}_j$ can be ``taken over'' by $\myvec{x}_k$. 

Thus, we implement a selective scaling strategy. While searching over $\beta_j$, we only allow the scaling of $\beta_k$ if the correlation between $\myvec{x}_j$ and $\myvec{x}_k$ is above some threshold, instead of scaling all $\myvec{\beta}_{-j}$. Let $E_j$ denote the set of variables that are sufficiently correlated with $\myvec{x}_j$, i.e.,
\begin{eqnarray}
E_j &=& \{k \neq j \mbox{ such that } |\mbox{corr}(\myvec{x}_k, \myvec{x}_j)| > \rho_{min}\}.\label{eq:inclset}
\end{eqnarray}
for some pre-chosen $\rho_{min}>0$. The selective scaling steps solve
\begin{multline}
\label{eq:MCPselect}
\underset{\beta_j,v}{\min}\quad
L(\beta_j,v) = 
\left\|\myvec{y} 
- \beta_j \myvec{x}_j 
- \sum_{\ell \in E_j^C\backslash\{j\}} \beta_{\ell}\myvec{x}_{\ell}
- \sum_{k \in E_j} (v\beta_{k})\myvec{x}_{k}
\right\|^2 + \\
J(\beta_j) + 
\sum_{\ell \in E_j^C\backslash\{j\}} J(\beta_{\ell}) + 
\sum_{k \in E_j} J\left(v\beta_{k}\right),
\end{multline}
for $j=1,2,...,d$. As mentioned above, we can compute the optimal $\beta_j$ and $v$ explicitly for this problem (see Appendix~\ref{sec:MCPdetail}).
\end{subsection}

\begin{subsection}{Fitting entire regularization surfaces with multiple warm starts}

Using the MC+ penalty (\ref{eq:mcp}), the optimal solution $\widehat{\myvec{\beta}}$ to (\ref{eq:PR}) depends on two regularization parameters, $\lambda$ and $\gamma$. For different values of $(\lambda,\gamma)$, one can think of $\widehat{\myvec{\beta}}(\lambda,\gamma)$ as tracing out an entire regularization surface. \citet{sparsenet} provided a nice algorithm, called SparseNet, for fitting the entire regularization surface. 

Fitting an entire surface of solutions, rather than just a single solution, introduces an interesting set of challenges for our work. When fitting a single solution, we are only concerned with how to best find a good solution for a given pair of $(\lambda,\gamma)$. However, SparseNet fits the entire surface of solutions sequentially, using each point on the surface, $\widehat{\myvec{\beta}}(\lambda,\gamma)$, as a warm start for fitting the ``next'' point. 
Thus, improving the solution at $(\lambda,\gamma)$ \emph{may} not be desirable if the improved solution provides an inferior warm start for 
the next point, resulting in worse solutions further down the surface. Empirically, we have found this to be a common occurrence. 

To remedy this problem, our strategy is to keep track of a few different solution surfaces:
\begin{itemize} 

\item $\widehat{\myvec{\beta}}_A(\lambda,\gamma)$ --- this is the ``usual'' surface obtained by SparseNet, i.e., each point on this surface is obtained by running the coordinate descent algorithm (an ultimate AO strategy), using the ``previous'' point on this surface (A) as a warm start;

\item $\widehat{\myvec{\beta}}_B(\lambda,\gamma)$ --- each point on this surface is obtained by first running the coordinate descent algorithm and then switching over to search in a different space, but each point also uses the ``previous'' point on this surface (B) as a warm start;

\item $\widehat{\myvec{\beta}}_C(\lambda,\gamma)$ --- like surface B above, each point on this surface also is obtained by first running the coordinate descent algorithm and then switching over to search in a different space, {\em except} that each point uses the ``previous'' point from the surface $\widehat{\myvec{\beta}}_A(\lambda,\gamma)$ as a warm start. 
\end{itemize}
At each point $(\lambda,\gamma)$, we keep the better of 
$\widehat{\myvec{\beta}}_B(\lambda,\gamma)$ or
$\widehat{\myvec{\beta}}_C(\lambda,\gamma)$ as our solution. 

\paragraph{Remark} In the actual implementation, it is clear that we need not start from scratch in order to obtain the surface $\widehat{\myvec{\beta}}_C(\lambda,\gamma)$; we can simply start with the surface $\widehat{\myvec{\beta}}_A(\lambda,\gamma)$, and apply our escaping strategies directly at each point. Conceptually, we think it is easier for the reader to grasp what we are doing if we describe three separate surfaces rather than two, but this does \emph{not} mean we have to triple the amount of computation. 

\end{subsection}

\end{section}

\begin{section}{Experimental results}

We now present some experimental results to demonstrate the effectiveness of our method. For matrix factorization, we use a real-world data set; for MC+ regression, we use a simulated data set. 

\begin{subsection}{Matrix factorization}
\label{sec:rsltMF}

To demonstrate our method for matrix factorization, we used a data set compiled by \cite{amazon-data}, which consists of approximately 35.3 million reviews from \url{www.amazon.com} between 1995 and 2013. We took a dense subset of their data consisting of approximately 5.5 million reviews, such that all users in our subset have rated at least 55 items, and all items have been rated at least 24 times. 

For our restricted joint optimization approach, we allowed only a small number ($s \ll n,m$) of user- and item-vectors in each round to participate in the joint optimization (see Section~\ref{sec:MFjoint}). Empirically, we obtained reasonably good and comparable performance results with a wide range of $s \in [20,200]$, but results reported here are for $s=50$.

We tested our method by randomly splitting the ratings into two halves, using one half as the training set $T$, and the other half as the test set $V$. All statistics were averaged over ten runs. As our metric, we used the mean absolute error (MAE),
\begin{eqnarray}
\text{MAE} = \frac{1}{|V|} \sum_{(u,i) \in V} |\widehat{r}_{ui} - r_{ui}|.
\end{eqnarray}
\citet{amazon-data} reported a mean squared error (MSE) of about $1.42$ on their full Amazon data set using the baseline matrix factorization method with $K=5$. This would translate to about $1.19$ on the root mean squared error (RMSE) scale, which is more comparable with the MAE. Here, our baseline AO produced slightly better results (see Table~\ref{tab2}) because we used a dense subset, so there is presumably more information to be learned about each user and item in our subset. 

In order to produce fair comparisons between different methods, for given $K=5, 10$ and $15$ we used cross-validation to choose an optimal value of $\lambda$ for each method. The optimal $\lambda$ values are shown in Table~\ref{tab1}, with the corresponding average MAEs shown in Table~\ref{tab2}. As can be seen in Table~\ref{tab2}, our approach produces meaningfully better models.

\begin{table}
\centering
\caption{\label{tab1}%
Matrix factorization example. Optimal $\lambda$'s chosen by cross-validation.}
\fbox{%
\begin{tabular}{l|cccc}
 & Baseline & Scaling & Random   & Greedy  \\ 
 & AO       & Only    & Subspace & Subspace  \\ 
 & (Sec.~\ref{sec:MF}) & (Sec.~\ref{sec:MFscale}) & (Sec.~\ref{sec:MFrand}) & (Sec.~\ref{sec:MFopt}) \\
\hline 
K=5  & 1  & 5 & 5  & 5  \\
K=10 & 9  & 5 & 9  & 9  \\
K=15 & 15 & 5 & 12 & 12 \\
\end{tabular}}
\end{table}

\begin{table}
\centering
\caption{\label{tab2}%
Matrix factorization example. Mean absolute error on the test set, $V$. }
\fbox{%
\begin{tabular}{l|cccc}
 & Baseline & Scaling & Random   & Greedy  \\ 
 & AO       & Only    & Subspace & Subspace  \\ 
 & (Sec.~\ref{sec:MF}) & (Sec.~\ref{sec:MFscale}) & (Sec.~\ref{sec:MFrand}) & (Sec.~\ref{sec:MFopt}) \\
\hline 
K = 5  & 0.856 & 0.763 & 0.747 & 0.740 \\ 
K = 10 & 0.859 & 0.756 & 0.760 & 0.754 \\ 
K = 15 & 0.861 & 0.760 & 0.769 & 0.760 
\end{tabular}}
\end{table} 

Figure~\ref{fig:MFrslt} shows that, while there appeared to be little difference (Table~\ref{tab2}) between using a random choice and using a greedy choice of $\{\myvec{w}_p^u, \myvec{w}_q^i\}$ to conduct the restricted joint search, the greedy strategy was much faster and more efficient at improving the results.

\begin{figure}[hptb]
\centering
\includegraphics[width=\textwidth]{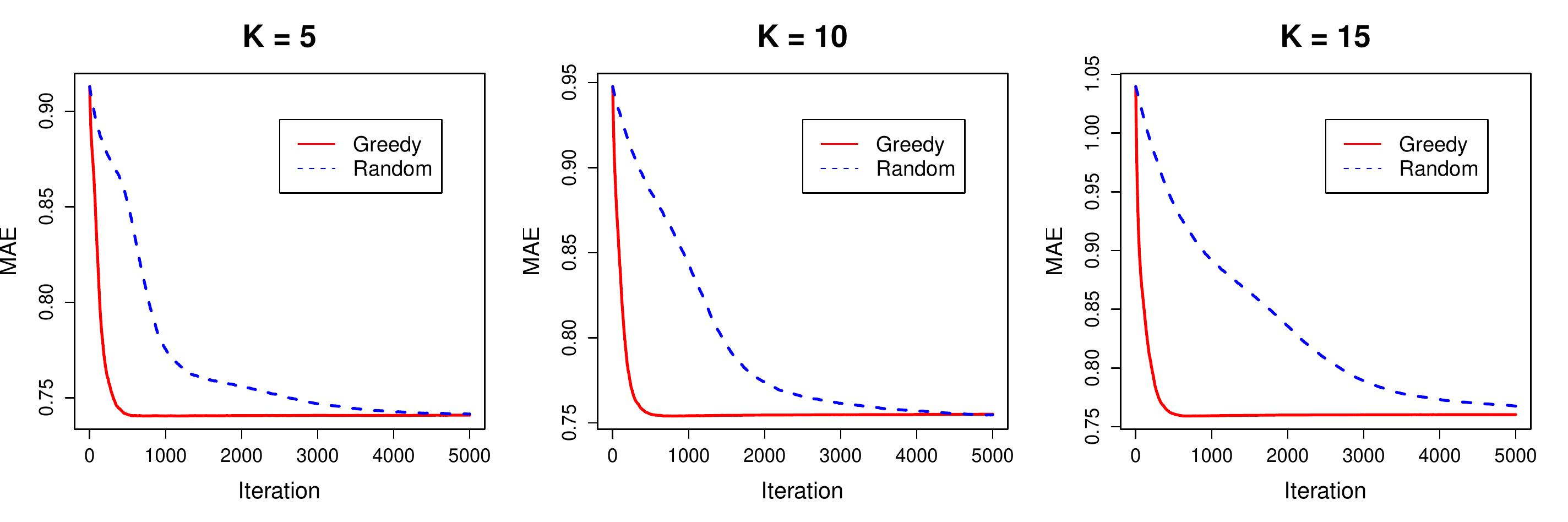} 
\caption{\label{fig:MFrslt}%
Matrix factorization example. The 
MAE on the test data versus the number of iterations.}
\end{figure}

\end{subsection}

\begin{subsection}{MC+ regression}
\label{sec:rsltMCP}

To demonstrate our method for MC+ regression, we used a simulated data set from \cite{sparsenet} --- more specifically, their model $M_1$. The sample size is $n=100$, with $d=200$ predictors generated from the Gaussian distribution with mean zero and covariance matrix $\myvec{\Sigma}$, whose $(j,k)$-th entry is equal to $0.7^{|j-k|}$. The response is generated as a linear function of only $10$ of the $200$ predictors plus a random noise; in particular, 
\[
\myvec{y} = \myvec{x}_1 + \myvec{x}_{21} + \myvec{x}_{41} + ... + \myvec{x}_{161} + \myvec{x}_{181} + \myvec{\varepsilon}.
\]  
That is, $\beta_{20j+1}=1$ for $j=0,1,2,...,9$ and $\beta_j=0$ otherwise. \citet{sparsenet} took $\varepsilon_i \sim N(0,\sigma^2)$ with $\sigma=\sqrt{\myvec{\beta}^{\T} \myvec{\Sigma} \myvec{\beta}}/3$ so that the signal-to-noise ratio is $3$.

Using $\rho_{min}=0.3$ in equation~(\ref{eq:inclset}), we estimated $\widehat{\myvec{\beta}}(\lambda,\gamma)$ on a grid consisting of $8$ different $\gamma$'s and $50$ different $\lambda$'s.
The $\gamma$'s were equally spaced on the logarithmic scale between $\gamma=1.000001$ and $\gamma=150$.
The $\lambda$'s were equally spaced on the logarithmic scale between $\lambda=\lambda_{max}$, which is the smallest $\lambda$ such that $\widehat{\beta}_j=0$ for all $j$, and $\lambda=0.01\lambda_{max}$. 

For each point on the grid we considered, we computed the percent decrease in the value of the objective function, i.e.,
\[
 \%\Delta_L = \frac{L_{new}-L_{old}}{L_{old}},
\]
where $L_{old}, L_{new}$ are the values of the objective function when the coordinate descent algorithm converged, and after our restricted joint search, respectively. 
Table~\ref{tab3} shows that, for about $72\%$ of points on the grid, our strategy made little difference ($\%\Delta_L$ no more than $0.5$ percentage points), indicating that the original coordinate descent algorithm already found relatively good solutions at those points. For the remaining $28\%$ of the points, however, our strategy found a better solution --- searching in a slightly expanded space further reduced the value of the objective function by an average of $5\%$. For the smaller half of $\gamma$'s (more nonconvex objective functions), the average percent decrease was a little over $6\%$; for the larger half (less nonconvex objective functions), the average percent decrease was close to $4\%$. 

For each point on the grid, we also computed the percent decrease in the variable-selection error, i.e.,
\[
 \%\Delta_e = \frac{\mbox{error}_{new}-\mbox{error}_{old}}{\mbox{error}_{old}},
\]  
where $\mbox{error}_{old}, \mbox{error}_{new}$ are the errors of the coordinate descent solution and of our solution, respectively. The variable-selection error was measured in terms of 
\begin{eqnarray}
\label{eq:varselerr}
\mbox{error}(\widehat{\myvec{\beta}}) &=& 
\frac{1}{d} \sum_{j=1}^d I(\beta_j=0 \mbox{ and } \widehat{\beta}_j \neq 0) + I(\beta_j \neq 0 \mbox{ and } \widehat{\beta}_j = 0).
\end{eqnarray}
Table~\ref{tab4} shows that, overall, our strategy led to improved variable-selection results as well --- a $2\%$ reduction in error on average. 

\begin{table}
\centering
\caption{\label{tab3}%
MC+ regression example. Comparison in terms of the terminal values of the loss function, $L$.}
\fbox{%
\begin{tabular}{l|rrr}
 & \multicolumn{1}{c}{Small $\gamma$'s} & \multicolumn{1}{c}{Large $\gamma$'s} & \multicolumn{1}{c}{All $\gamma$'s}\\
\hline 
$\mbox{fraction}(-0.005\leq\%\Delta_L<0)$         &  $0.715$ &  $0.730$ &  $0.723$\\
$\mbox{average}(\%\Delta_L|\%\Delta_L<-0.005)$    & $-0.063$ & $-0.036$ & $-0.050$ 
\end{tabular}}
\end{table}

\begin{table}
\centering
\caption{\label{tab4}%
MC+ regression example. Comparison in terms of the variable-selection error (\ref{eq:varselerr}).}
\fbox{%
\begin{tabular}{l|rrr}
 & \multicolumn{1}{c}{Small $\gamma$'s} & \multicolumn{1}{c}{Large $\gamma$'s} & \multicolumn{1}{c}{All $\gamma$'s} \\
\hline 
$\mbox{fraction}(\%\Delta_e=0)$                   &  $0.575$ &  $0.205$ &  $0.390$  \\
$\mbox{average}(\%\Delta_e|\%\Delta_e \neq 0)$    & $-0.011$ & $-0.027$ & $-0.021$
\end{tabular}}
\end{table}

\end{subsection}

\end{section}

\begin{section}{Conclusion}
In this article, we have proposed a general framework for improving upon alternating optimization of nonconvex functions. The main idea is that, once standard AO has converged, we switch to conduct our search in a different subspace. We have provided general guidelines for how these different subspaces can be defined, as well as illustrated with two concrete statistical problems --- namely, matrix factorization and regression with the MC+ penalty --- how problem-specific information can (and should) be used to help us identify good and meaningful search subspaces. By carefully selecting a relevant space to search over, we can escape undesirable locations such as saddle points and produce notable improvements. In addition to serving as examples of our general idea, we think that these improved AO algorithms, for matrix factorization and for regression with the MC+ penalty, are meaningful contributions on their own.  
\end{section}

\section*{Acknowledgments}
This research is partially supported by the Natural Sciences and Engineering Research Council (NSERC) of Canada and by the University of Waterloo.

\appendix
\begin{section}{Explicit solution of problem (\ref{eq:MCPselect})}
\label{sec:MCPdetail}

This appendix gives details of how the problem (\ref{eq:MCPselect}) can be solved explicitly. This is done by identifying all points that satisfy the first order conditions, as well as those at which the objective function $L(\beta_j, v)$ is not differentiable, and choosing from all these points the one that minimizes the objective function. 

\subsection{First order conditions}

First, we define two (vector) constants,
\begin{eqnarray}
\myvec{c} = \sum_{k \in E_j} \beta_k \myvec{x}_{k}, \quad
\myvec{d} = \sum_{\ell \in E_j^C\backslash\{j\}} \beta_{\ell} \myvec{x}_{\ell} 
\end{eqnarray}
and re-write the objective function (\ref{eq:MCPselect}) as
\begin{align}
\label{eq:MCPselect2}
L(\beta_j,v)& = 
\frac{1}{2} \left\| \myvec{y} - \beta_j \myvec{x}_{j} - v \myvec{c} - \myvec{d} \right\|^2 + 
J(\beta_j) + 
\sum_{k \in E_j} J(v \beta_k) +
\sum_{\ell \in E_j^C\backslash\{j\}} J(\beta_{\ell}), 
\end{align}
where $J(\cdot)$ is the MC+ penalty function given by (\ref{eq:mcp}). The (vector) constants $\myvec{c}$ and $\myvec{d}$ are terms that do not depend on either $\beta_j$ or $v$. The first-order conditions are then given by
\begin{align}
\label{eq:1stbeta}
\frac{\partial L}{\partial \beta_j} &= 
 - \myvec{x}_{j}^{\T} (\myvec{y} - \beta_j \myvec{x}_j - v \myvec{c} - \myvec{d}) + 
 J^{\prime}(\beta_j)=0 
\end{align}
and
\begin{align}
\label{eq:1stv}
\frac{\partial L}{\partial v} &= - \myvec{c}^{\T}(\myvec{y} - \beta_j \myvec{x}_j - v \myvec{c} - \myvec{d}) + 
 \sum_{k \in E_j} \beta_k J^{\prime}(v\beta_k) = 0,
\end{align}
where $J(t)$ is not differentiable at $0$, and for $t \neq 0$,
\begin{align}
J^{\prime}(t) = 
\begin{cases}
\lambda \left[\text{sgn}(t)\right] - \frac{|t|}{\gamma}, &  |t| \leq \gamma \lambda; \\
0,                                                     &  |t| > \gamma \lambda.
\end{cases}
\end{align}

\subsection{Expression of $\beta_j$ for fixed $v$ }
\label{appdx:beta}

Recall that we assume $\|\myvec{x}_j\|=1$ for all $j$ (Section~\ref{sec:PR}). For given $v$, then, equation~(\ref{eq:1stbeta}) implies that any solution $\beta_j \neq 0$ can be expressed as
\begin{align}
\label{eq:1stBans}
\beta_j = 
  \underbrace{\left[\frac{\myvec{x}_{j}^{\T}(\myvec{y} - \myvec{d}) - C}{1 - r}\right]}_{\psi} + 
v \underbrace{\left[\frac{- \myvec{x}_j^{\T}\myvec{c}}{1 - r}\right]}_{\xi} 
\equiv \psi + \xi v,
\end{align}
where $C$ and $r$ (and hence $\psi$ and $\xi$ as well) are different constants depending on whether $|\beta_j| > \gamma\lambda$ and whether $\beta_j$ is positive or negative. In particular,  
\begin{align*}
C = \begin{cases}
0,                                        & \text{if } |\beta_j| > \gamma \lambda; \\
\lambda \left[\text{sgn}(\beta_j)\right], & \text{if } |\beta_j| \leq \gamma \lambda \mbox{ and } \beta_j \neq 0
\end{cases} 
\end{align*}
and
\begin{align*}
r = \begin{cases}
0,                                       & \text{if } |\beta_j| > \gamma \lambda; \\
\left[\text{sgn}(\beta_j)\right]/\gamma, & \text{if } |\beta_j| \leq \gamma \lambda \mbox{ and } \beta_j \neq 0.
\end{cases}
\end{align*}
Without knowing where $\beta_j$ is a priori, our strategy is to proceed with solving for $v$ (Section~\ref{appdx:v} below) using different $(\psi,\xi)$-pairs, and discarding ``solutions'' that turn out to be inconsistent. For example, if a particular ``solution'' $\beta_j$ is obtained using a $(\psi,\xi)$-pair that assumes $-\gamma\lambda \leq \beta_j <0$ --- i.e., $C=-\lambda$ and $r=-1/\gamma$ in (\ref{eq:1stBans}) --- but $\beta_j$ turns out to be outside this interval, such a ``solution'' is automatically discarded. 

For $\beta_j=0$, we proceed to solve for $v$ by setting $(\psi,\xi)=(0,0)$. If multiple solutions exist that consistently satisfy the first order conditions, then the one that minimizes (\ref{eq:MCPselect2}) is chosen.

\subsection{Solving for $v$}
\label{appdx:v}

For any given $(\psi,\xi)$-pair --- including $(\psi,\xi)=(0,0)$, substituting (\ref{eq:1stBans}) into equation~(\ref{eq:1stv}) gives 
\begin{align}
- \myvec{c}^{\T}(\myvec{y} - \psi \myvec{x}_j - v(\myvec{c} + \xi \myvec{x}_{j}) - \myvec{d}) + 
\sum_{k \in E_j} \beta_k J^\prime(v\beta_k) = 0.
\end{align}
This is a single, piecewise-linear equation in a single variable $v$, which can be solved explicitly. First, we check whether $v=0$ is a solution. Then, we consider the cases of $v>0$ and $v<0$ separately.  

\subsubsection*{The case of $v > 0$}

When $v > 0$, we have 
\begin{align*}
J^\prime(v \beta_k) = 0 \quad\Longleftrightarrow\quad |v \beta_k| > \gamma \lambda \quad\Longleftrightarrow\quad v > \frac{\gamma \lambda}{|\beta_k|}.
\end{align*}
Let $K=|E_j|$ be size of the set $E_j$.
For all $\{\beta_k: k \in E_j\}$, we use the notation $\beta_{(k)}$ to indicate that $\beta_{(k)}$ is $k$-th smallest member of the set {\em in terms of its absolute value}. That is,
\[
|\beta_{(1)}| \leq \cdots \leq |\beta_{(k-1)}| \leq |\beta_{(k)}| \leq \cdots \leq |\beta_{(K)}|.
\]
We then define a partition as follows: 
\begin{eqnarray*}
  I_{K} &=& \left[0,\frac{\gamma \lambda}{|\beta_{(K)}|}\right), \\
  I_{k-1} &=& \left[\frac{\gamma \lambda}{|\beta_{(k)}|},\frac{\gamma \lambda}{|\beta_{(k-1)}|}\right) \quad \mbox{for all}\quad K \geq k > 1,\\ 
  I_0 &=& \left[\frac{\gamma \lambda}{|\beta_{(1)}|},\infty\right).
\end{eqnarray*}
On each interval $I_{k-1}$, we have not only $v \geq \frac{\gamma \lambda}{|\beta_{(k)}|}$, but also $v \geq \frac{\gamma \lambda}{|\beta_{(k')}|}$ for all $k' \geq k$, which means $J^\prime(v \beta_{k'}) = 0$ for all $k' \geq k$. 
Thus, on a given interval $I_{k-1}$, equation (\ref{eq:1stv}) becomes
\begin{align}
\label{eq:linear-v}
-\myvec{c}^{\T}
\left[\myvec{y} - \psi \myvec{x}_{j} - v(\myvec{c} + \xi \myvec{x}_{j}) - \myvec{d}\right] + 
\sum_{k' < k} \beta_{k'} \left(\lambda \left[\text{sgn}(\beta_{k'})\right] - \frac{v \beta_{k'}}{\gamma}\right)= 0.
\end{align}
Notice that $\text{sgn}(v\beta_k)=\text{sgn}(\beta_k)$ if $v > 0$.
Since equation~(\ref{eq:linear-v}) is linear in $v$, to check for its zeros, it suffices to evaluate the left-hand side at the endpoints of $I_{k-1}$ and determine if there is a change of sign. If there is, we can solve for $v$ as 
\begin{align}
\label{eq:v}
v = \frac{ \myvec{c}^{\T}(\myvec{y} - \psi \myvec{x}_{j} - \myvec{d}) - 
           \lambda \sum_{k' < k } \beta_{k'} \left[\text{sgn}(\beta_{k'})\right]}
         { \myvec{c}^{\T}(\myvec{c} + \xi \myvec{x}_{j}) - 
		   \sum_{k' < k} \frac{\beta_{k'}^2}{\gamma}}. 
\end{align}
Notice that equation~(\ref{eq:linear-v}) may have solutions on multiple intervals, that is, for more than one $k$. Each solution $v$ will lead to a different solution for $\beta_j$ --- see equation~(\ref{eq:1stBans}). As we already pointed out in Section~(\ref{appdx:beta}), some of these ``solutions'' may be inconsistent and discarded accordingly; if more than one consistent solutions remain, then the one that minimizes (\ref{eq:MCPselect2}) is kept.

\subsubsection*{The case of $v < 0$}

When $v < 0$, we can perform the exact same search, except that each interval $I_k$ is now  the reflected about $0$, and that the term $\text{sgn}(\beta_{k'})$ in (\ref{eq:v}) is multiplied by $-1$ as $\text{sgn}(v\beta_k)= -\text{sgn}(\beta_k)$ if $v < 0$. 

\end{section}

\bibliographystyle{/u/m3zhu/natbib}
\bibliography{./references}
\end{document}